# A Grid Based Architecture for High-Performance NLP [1]

**Baden Hughes and Steven Bird**
Department of Computer Science and Software Engineering
University of Melbourne
Victoria, 3010, Australia
`{badenh, sb}@cs.mu.oz.au`

## Abstract

We describe the design and early implementation of an extensible, component-based software architecture for natural language engineering applications which interfaces with high performance distributed computing services. The architecture leverages existing linguistic resource description and discovery mechanisms based on metadata descriptions, combining these in a compatible fashion with other software definition abstractions. Within this architecture, application design is highly flexible, allowing disparate components to be combined to suit the overall application functionality, and formally described independently of processing concerns. An application specification language provides abstraction from the programming environment and allows ease of interface with high performance computational grids via a broker.

## 1   Introduction

Computational grids are an emerging infrastructure framework for conducting research where problems are often data- or processor-intensive. A computational grid allows for large-scale analysis, distributed resources and processing, in addition to engendering new models for collaboration and application development. Given the prevalence of large data sources in the natural language engineering domain and the need for raw computational power in the automated analysis of such data, the grid computing paradigm provides efficiencies otherwise unavailable to natural language engineering.

Language engineering applications are typically constructed from a number of processing components, each responsible for a specialized task. Typical components include speech recognition, tagging, entity detection, anaphora resolution, parsing, etc. Each component is heavily parameterized and must be trained on very large datasets (e.g. the LDC Gigaword corpus (Graff, 2002)). Discovering optimal parameterizations is both data- and processor-intensive. Building complex applications, such as spoken dialogue systems, depends on identifying and integrating suitable components often from a range of sources.

In this paper we describe the design and early implementation of an extensible, component-based software architecture for natural language engineering applications which allows interfaces with high performance computing services. Recent advances in the automated exposure and harvesting of language resource catalogues are extended for the natural language processing domain. Based on these extensions we can programmatically and intelligently discover data sources, components, applications and computational grid nodes offering specific services. In such a model, a series of distributed individual components are assembled via an application framework, with internal communication requirements addressed by a common interface specification. An application consists of one or more data sources, together with a

---

number of processing components, coordinated within an overall framework. Components are function-specific implementations which adhere to a core series of standards for inter-component coordination which is open to extension by third parties.

Collections of components which form applications are then expressed using an application specification language which is parsed by a broker that interfaces with the computational grid infrastructure. Any application composed within or outside this model can address such a broker, and hence access the power and scalability of computational grids for processing. Furthermore, resultant data can be stored for future re-use either as static data, or as a new data-based service.

This model has numerous benefits. First, existing natural language engineering applications can benefit from grid services without being grid-aware, simply by instructing the broker to perform a specified task as if it were a single server. Second, application definitions are a declarative specification of a processing task and of the relationship between processing components, facilitating substitution of equivalent components and aggregation of simple applications into more complex applications. Third, application definitions and result sets can be stored, described using standard metadata, and be discovered and re-used by other applications at a later date.

In this paper, we discuss the architectural foundations of our proposal, resource discovery mechanisms, component identification, multi-component application design (including several sample applications), the grid services interface and the application specification language. Furthermore some directions for future work are identified.

## 2 Proposed Architectural Model

The architecture we propose has four discrete units, namely: 1) *data*, which we consider to be language resources that are potentially destined for analysis; 2) *components*, which we define as minimal individual functional units, all of which subscribe to a common interface specification; 3) *applications*, which we consider to be an aggregate representation of data, components and a logical execution order; and finally 4) *grid services*, which are network accessible computational or storage resources on which an application can be executed or reside as a service.

The processing model we envisage allows the free-standing description of both data and components in a manner compatible with existing metadata standards. These descriptions are aggregated and indexed, and able to be queried via a resource discovery agent or used as data sources eg through an application design portal. After data and components are selected, they are then combined into a single description (an XML document), which can be parsed by a broker which interfaces with grid services to execute the required tasks, taking into account task completion preferences such as deadlines, available services, economy and bandwidth. An overview of this model is displayed below in Figure 1.

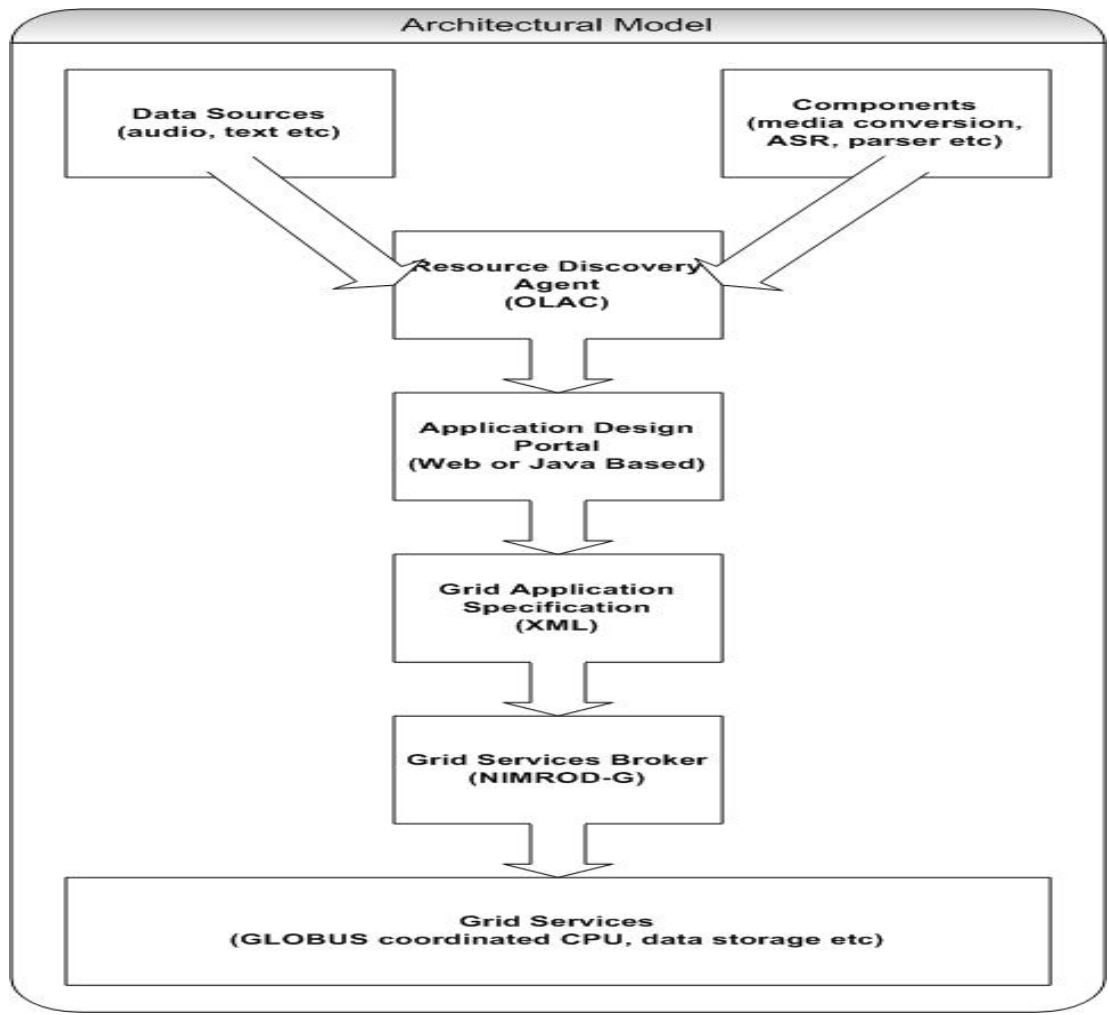

**Figure 1. Architectural Model**

In proposing this architecture and its constituents, we are conscious of synergistic efforts within the NLP domain in addressing these high level requirements. A number of researchers are beginning to explore the interface between high performance distributed computing and natural language processing, we discuss briefly two recent proposals which are broadly aligned with our own.

Curran (2003) provides a useful insight into infrastructure requirements in his consideration of how to integrate compiled applications for high performance computing environments. Curran primarily describes integration tasks as being focused on optimization of coding and compilation as a means to gain accuracy, speed and compactness, but also identifies two aspects of interest to ourselves – namely resource re-use at the level of both input and output; and descriptive abstractions – although motivated by the web services approach of providing function names and arguments over a lower level protocol.

Krieger (2003) also identifies the need for a mechanism to connect individually developed NLP components through the use of wrappers to expose a pseudo-API in the absence of formal API documentation. Krieger advocates the use of typed feature structures to describe functionality, which in part is analogous to our approach. However, we motivate functional description from the domain of metadata for resource discovery, and combine this formalism with other general approaches to the description of processing tasks.

## 2.1 Grid as Infrastructure

As noted earlier, computational grids are an emerging infrastructure framework for research in high performance computing. A computational grid typically consists of a number of distributed machines which are coordinated centrally, and provide processing or data storage services.

A computational grid also allows parallel and parametric processes to be executed across a number of processing nodes, providing scalable, efficient and robust computational facilities. The potential of grid computing is of interest in the natural language engineering domain where applications are typically constructed out of numerous components (each responsible for a specialised task), and executed against large data sets. Applications of this order in other domains (e.g. meteorology – for climate modeling, biochemistry – for protein analysis, physics – for particle and fluid dynamics) are already taking advantage of the computational power provided by the grid computing paradigm.

In order to leverage computational grids for natural language engineering research, we need to subscribe to an architectural model which allows automated discovery of data and components, flexible design of applications, coordination of execution, and storage of results. Preferably, such a model will easily allow researchers to design their applications and execute these over the computational grid without needing to acquire expertise in grid computing itself, which we view as a infrastructural framework which can be commoditized.

For further details on grids, Foster et al, (2001, 2002) provides a physiological and an anatomical overview of grid computing services, and provides foundational architectures for application development in the grid space.

## 2.2 Basic Processing Paradigms

A range of software architectures for natural language engineering have been developed in the last decade, and matured to the extent that there are a wide range of applications built on foundations such as GATE (Cunningham et al, 1996, Cunningham 2002), and ATLAS (Bird et al, 2000). However, the majority of these software architectures envisage a localized processing model, where data sources, entire applications, and computational resources are co-located. (One notable exception is the Annotation Graph Toolkit, which envisaged a database server model which could be remote – see Cieri and Bird 2001, Ma et al 2002.) In addition, the mode of processing is generally assumed to be serialized, with little concern for the efficiencies which can be gained from parallel and parametric processing, nor the advantages of simultaneous processes being interoperable.

Equally, there have been numerous developments for the deployment of computational grids, ranging from middleware, API's, libraries and infrastructural management approaches. Within grid architectures currently deployed, there are both processor-centric and data-centric architectures, of which the most common is the processor-centric grid. In this context, grid-aware applications typically adopt the approach of transporting the data from a storage site to the location of available CPU capacity.

Skillicorn (2001) identifies a number of computational grid types, of which the most common is the processor-centric grid as referred to earlier. In natural language engineering this model is less attractive because the size of data sources is typically significant enough for any network based transport to be relatively inefficient and cost-prohibitive. Skillicorn advocates the concept of the data-centric computational grid, where grid-aware applications are transported to the location of the data source, and describes a number of implications for architectural models relevant to natural language engineering. In this area, the requirements of natural language engineering may provide the catalyst for such an innovative broker design.

The model advocated in this paper would work equally well in either context providing that a sufficiently intelligent broker architecture was available. At the time of writing, it is envisaged that a processor-centric architecture will be utilized at least in the medium term whilst broker implementations are upgraded to support the data-centric model. Additionally, a number of evolutionary steps in network hardware and software design are required to take full advantage of a data-centric processing model.

## 2.3 Resource Discovery

Within any distributed natural language engineering system, there clearly exists a need to discover data sources, components, applications and processing services over the network. This need is expressed both by the application developer, who desires to build an application from a range of available components, as well as by the brokering agent, which needs to align application requirements with available computational resources.

There are a range of methods available to facilitate such discovery. Of note are the potential for the use of a Web Services implementation (using DAML-S (DAML Services Coalition, 2002) or RDF (Lassila and Swick (eds.) 1999) ), or embedding this task within the wider Semantic Web implementations (e.g. WSDL (Chinnici, et al 2003)). Whilst the authors are appreciative of the developments within these areas, there already exists a language resource specific framework for resource description and discovery within the Open Language Archives Community (OLAC) which has significant benefit, and it is on this foundation that we have built.

The Open Language Archives Community (OLAC) Metadata Set (Bird and Simons, 2002a) is an emerging metadata standard for description of language resources, including software. More than 30,000 language resources from over 20 archives are described using OLAC metadata. Objects described using OLAC can be harvested and aggregated through the use of the Open Archives Initiative Protocol for Metadata Harvesting (OAI-PMH) (Lagoze et al 2002). The resulting centrally-stored index can be queried to locate objects of interest. An advantage of this approach is that the OLAC Metadata is an extension of the Dublin Core Metadata Set (Dublin Core Metadata Initiative, 2003), thus ensuring widespread accessibility. The OLAC initiative has a standardized mechanism for extending OLAC metadata (Bird and Simons, 2002b) which we have adopted to describe data sources, applications and processing nodes.

We have extended the OLAC resource description and discovery mechanisms by using a formal XML Schema extension. This work has adopted and extended standards based on the foundational work of Hughes (2002) for the encoding of information for the language technology domain – specifically CPU, Operating System, Programming Language, Source Status, Media Type, and Licensing. Electing to extend this standard allows us to leverage discovery tools and techniques already in existence and considered the benchmark for such processes.

Having based resource descriptions on OLAC standards we can then query an OLAC aggregator to discover the existence and status of resources of interest. From this heritage we can directly support other data and service description standards such as RDF and Web Services (eg by extending WSDL through the use of RDF Schema), whilst describing objects which do not easily fit within their traditional domains. In addition, using the same metadata set we can describe new services which may result from data processing through a grid application.

## 3   Component Identification

In order to explore the grid computing paradigm through a natural language engineering framework, a component based model has been adopted. Within this model, we identify and describe a number of reusable component types which can be combined to create multi-component applications to be executed in the grid environment. Each of these component types has as a common functional core the ability to communicate with a central management infrastructure using standard messaging interfaces. Each component must therefore be extended to support this communication system, which could be done by linking to a shared abstraction layer. In developing prototype implementations, we intend to wrap existing components wherever possible.

### 3.1   Annotation Server

A large class of language engineering tasks involving time-series data can be construed as adding a new layer of annotation to existing data. For example, a process which takes speech input and produces text output (e.g. a speech recognizer) can be viewed as adding textual annotation to audio data, with the result that both data types remain accessible for further analysis. When several tasks operate in this mode, they collectively build a rich store of linguistic information. For example, a prosody recognition component could identify major phrase boundaries in spoken input, and a parser could then employ both the ASR and prosody output in constructing parse trees.

Annotation graphs (Bird and Liberman, 2001) are labeled, directed graphs having optional time stamps at the nodes. They can be used to represent a diverse range of time-series annotations, including ASR output, POS tags, named entities, syntactic chunks or trees, aligned translations, dialogue acts, and so on. Importantly, the intermediate stages of many processing tasks are well-formed as annotation graphs, so certain outputs can be made available shortly after processing starts, facilitating efficient pipe-lining and streaming. The Annotation Graph Toolkit (AGTK) (http://agtk.sf.net) is an open source im-plementation of annotation graphs which works with a mature API and a wide range of models and formats. The Annotation Server is a component that collects, collates and stores annotation graphs which may be generated or accessed by other components and applications and extends on proposals previously described by Cieri and Bird (2001), and available as a prototype in the current AGTK distribution.

### 3.2 Automatic Speech Recognition and Forced Alignment

An ASR component can be used to create transcript, again represented directly as an annotation graph. An Alignment component is used to forcibly align digitized speech with a supplied transcript. This can be performed at various levels of granularity, and partial results can be naturally represented as an annota-tion graph.

### 3.3 Data Source Packaging

A Data Source Packaging component divides data sources into logical units which can be distributed across the grid as individual processing tasks. As an example, consider a digital audio file of 500Mb in size. As this would be non-trivial to transfer to a remote processing node, a Packager could divide this data source into fifty 10Mb digital audio files, each of which could be compressed, transported and proc-essed separately.

### 3.4 Conversion

A Conversion component is used to convert between media formats, character encoding schemes, and annotation types. The expected input and output formats of existing natural language engineering com-ponents are often incompatible, and this component will facilitate component integration, simplifying the task of wrapping existing components. In the case of media formats, this component will be able to serve a static data source as a stream source and vice versa. In the case of annotation conversion, this compo-nent will be able to convert between the wide variety of existing time-series annotation formats under-stood by AGTK.

### 3.5 Text Annotation

A Text Annotation component augments existing annotated text with new layers of annotation, e.g. POS tags, sense tags, co-reference, named entities, syntactic chunks, morphological analyses, translations, summaries and so forth.

### 3.6 Lexicon Server

A Lexicon Server component aggregates and stores lexical data that is generated or accessed by other components. Examples include the TIMIT lexicon (words and pronunciations from the TIMIT database) and WordNet. In the absence of a universal lexicon API there will need to be a family of commonly sup-ported APIs. The Lexicon server could be viewed as a service, where multiple other application instances can execute queries against its database.

### 3.7 Semantic Mapping

A Semantic Mapping component constructs and collates content based on theme identification. Themes may be defined by a list of keywords of concordance based output. Alternatively the mapping component can map term instances to an ontology for the particular domain.

## 4 Formal Component Description

Having now identified a number of different components, we turn to a discussion of their formal description. Our formalization is expressed as an XML document, with all the inherent functionality of this standards family. We inherit various segments from a range of other sources, including functional standards from Hughes' functionality schema (Hughes, 2002), and the constraints (expressed as requirements) from Dublin Core's model. As a result of our wrapping approach we can simplify many components to support a basic I/O interface, rather than handling an extremely diverse range of possibilities.

In considering the XML document, a component has the following elements: identifier, functionality, requires, input type, and output type. The `identifier` is qualified by two attributes: uri, which indicates the location of the component either by a URL or a repository identifier; and name, which currently is an arbitrary string which describes the software component. The `functionality` element is qualified by a single attribute, type, which inherits its entries from the OLAC controlled vocabulary for the description of language technology functionality. The `requires` element has multiple attributes which build on the same controlled vocabularies as functionality, and allows expression of technical constraints of the particular software component e.g. CPU, OS, language, license). The `input` and `output` elements each are qualified by an attribute pair type and format (modeled after Internet MIME types), which allow specification of the modality and media type with which the component will engage. We provide an example description in Figure 2 below.

```
<component>
    <identifier
        uri="http://mywebserver.com/sph2pipe"
        name="sph2pipe"/>
    <functionality type="media_conversion"/>
    <requires
        cpu="x86"
        os="unix"
        sourcestatus="compiled"
        license="ldc"/>
    <input type="audio/wav"/>
    <output type="audio/sph"/>
</component>
```

**Figure 2. A Sample Component Description**

## 5 Multi-Component Application Design and Description

Having now identified, described and wrapped such components, we propose to combine them to create a range of multi-component applications which can be executed over the distributed grid infrastructure. It is envisaged that applications will be assembled using a portal based approach for ease of deployment. Based on the components described earlier, we can build a number of different applications and application types. Next we discuss two of these applications, namely spoken passage retrieval and collaborative annotation. In both of these application instances there is a significant requirement for data and processor-intensive computation.

## 5.1 Applications

The Spoken Passage Retrieval application will allow a user to identify passages of interest within spoken document collections, e.g. a database of broadcast news. This application requires accurately transcribed and indexed sources – implying a need for media conversion, automatic speech recognition, transcription, forced alignment, entity detection, and annotation indexing. The resulting index will be viewed as analogous to a lexicon, which permits lookup on words which can then be located within particular regions in document of interest. The preparatory processes can be executed in parallel on a grid architecture, leveraging greater computational power to assist in data intensive tasks such as real time transcription and language modeling, where results from newly analysed data may lead to classifications which need to be applied retrospectively to the entire existing collection.

The Collaborative Annotation application will allow a user to mark features within a given data set. There are four basic types of collaborative annotation, and the model proposed supports them all. First, peers may cooperate in the construction of a large annotated resource, dividing up the workload and each performing the same task on their parcel of work. Limited coordination is necessary in order to ensure that all items are annotated appropriately (e.g. that 10% of items are doubly annotated for reliability testing). In the second type of collaboration, a supervisor vets the work of hired annotators, and optionally adds further codes that depend on their own critical analysis of the data. Third, two specialists may annotate the same dataset according to different theoretical models, in order to investigate the empirical differences between the theories and the extent to which categories posited by one theory can be derived from the other. Finally, collaboration may take place between human and automatic agents, in which the automated agents monitors human decision and, over time, develop increasingly refined models that predict those decisions and expedite the work. The latter case is of significant interest in the grid context where automated agents may rapidly annotate large datasets using a computational grid, thus providing human annotators a series of structural hypotheses which can be assessed. In turn, automated agents may observe human annotation, and where necessary, retrospectively apply these annotations to existing data sources.

## 5.2 Formal Application Description

We now turn to the formal description of the components which together form an application. Again, our formalization is expressed as an XML document, with inheritance of segments from component descriptions as well as other XML-based architectural services.

In considering the XML document, an application consists of a one or mode data sources, one or more components, and a logical execution order. The `datasource` element(s) have 3 qualifying attributes – URI, format and language. The `component` element(s) are directly inherited from the formal component descriptions. The `pipeline` element is simplified dervation from the XML-Pipeline specification (Walsh & Maler (eds.) , 2002) and can inherit any, or all of the properties inherent in that specification, which together define the logical execution order.

We provide an example description in Figure 3 below.

```
<application>
    <datasource
        uri="http://mywebserver.com/data/sample.wav"
        format type="audio/wav"
        language="EN"
    />
<component>
    <identifier
        uri="http://mywebserver.com/sph2pipe"
        name="sph2pipe"/>
    <functionality type="media_conversion"/>
    <requires
        cpu="x86"
        os="unix"
        proglang="c"
        license="ldc"/>
    <input type="audio/wav"/>
    <output type="audio/sph"/>
</component>
<pipeline
    <process id="1" component="sph2pipe"/>
<pipeline/>
</application>
```

**Figure 3. A Sample Application Description.**

## 6   Interfaces to Grid Services

The core infrastructure requisite for grid services can be provided in a number of ways. Historically, bootstrapping approaches such as those exhibited by Parallel Virtual Machines (PVM) (Geist et al, 1994) and Message Passing Interface (MPI) (Message Passing Interface Forum, 2001) have gained widespread acceptance. Owing to the low level intergration of such methods within application frameworks, significant development effort is required for integration, in addition to the technical management overhead of network resources required to use such frameworks. As such, we view these approaches as strewn with impediments to widespread adoption within our community of focus.

Instead, we have adopted the approach provided by Globus (Foster and Kesselman, (1997); Globus Project, (2003)). The Globus model allows for a distributed set of resources (a *grid*), upon which applications can execute in parallel or parametrically. Furthermore, within the Globus model, there is capacity for lightweight, cross-platform implementations which are ideally suited to the natural language engineering context where applications are often tied to a particular platform or architecture or local dataset. Coordination of grid services is an included component within the Globus framework, through the notion of the virtual organization (VO), and accessible directly through a middleware layer or by broker services which manage processes within the VO. These concepts are illustrated in Figure 4 below.

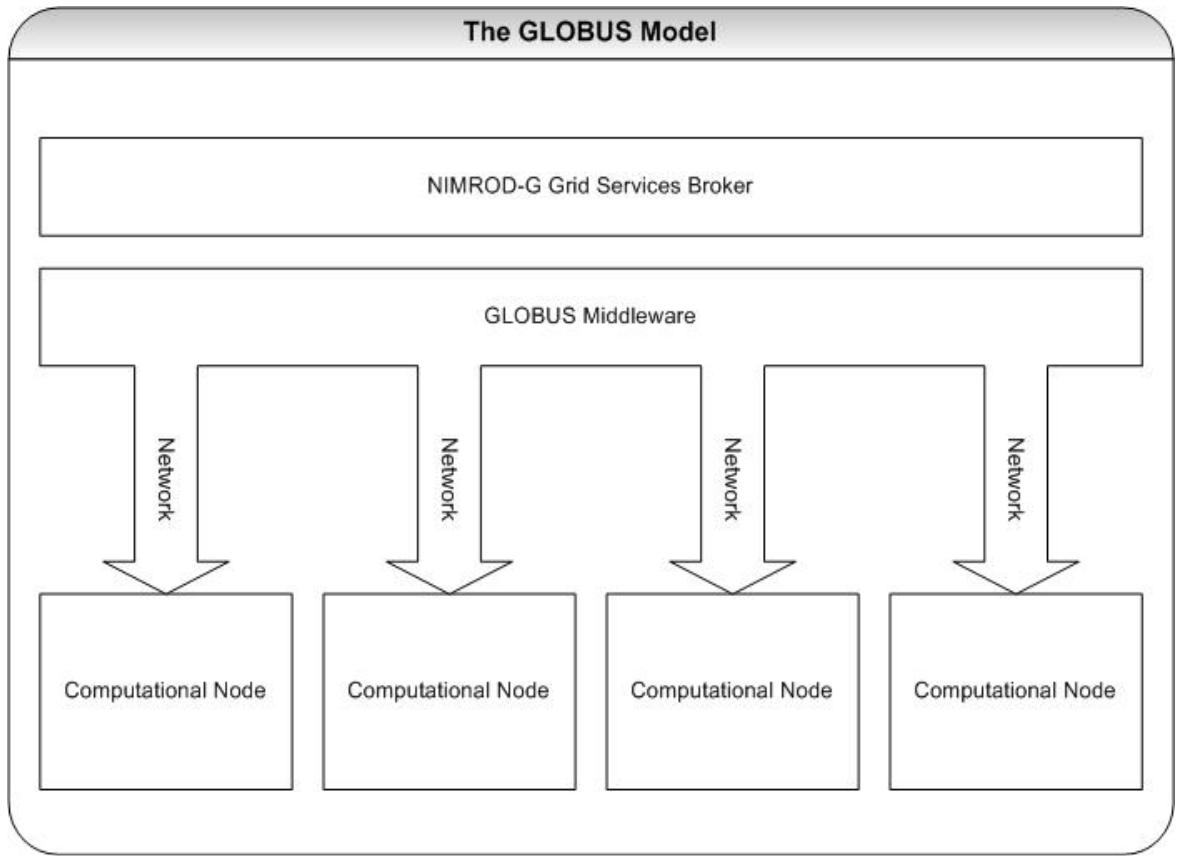

**Figure 4. The Globus Model**

The model of grid service interaction we have selected is a broker architecture which completely manages all aspects of grid interaction. This simplifies application development and leverages existing research in the area of broker architectures for grid environments. Within the grid environment, the NIMROD-G broker architecture (Buyya, et al, 2001) provides a well-tested, hardened, real-world broker architecture on which we can build, rather than attempting to building our own version from the ground up. The NIMROD-G architecture combines broker functions with grid services, for example, automated discovery of resources, as well as negotiation for services and dispatch management. In the context of this project, we have worked with the NIMROD-G developers to extend the NIMROD-G Resource Broker to parse the XML application specification language described earlier. This interface is independent of application framework and hence available to any NLE framework which requires access to grid-based services for processing tasks. In addition, services such as the Grid Market Directory (GMD) (Yu et al, 2002) have been extended to include support for requirements exhibited by NLP applications, but which are able to be generalized. Specifically, the GMD now supports a fine level of granularity for the description of grid node hardware and software constraints inherited from the semantics ascribed to our component descriptions.

## 7    Future Directions

Based on our work to this stage, we can divide our ongoing research into two categories, 1) formalism, and 2) architecture.

## 7.1    Formalism

In the domain of formalism, our proposal is based on the model specified in Buyya et al (2003), but is implemented in pure XML and inherits a number of features from metadata standards, particularly Dublin Core. We are extending the NIMROD-G broker architecture to accept input based on the natural language engineering application requirements, and will create the independent processing tasks, schedule them for processing on the grid, facilitate inter-task coordination and collate the results, storing them in a repository for later re-use as either data sets or services. Our Application Specification Language will allow interaction with a NIMROD-G application prototype and includes the ability to substitute variables both statically (at commencement) and dynamically (as the result of other tasks).

The formal specification language will support the full range of semantics to adequately describe data, components, applications, processing logic, communication frameworks and their aggregates. We envisage that data requirements include the location of static corpora and the interfaces for "dynamic" corpora (those accessible via a public API either provided externally or as a result of earlier processing), and any requirements for media format, encoding scheme or annotation type which can be selected from or derived from the corpora. Processing requirements include the CPU, memory, storage and any task completion deadlines. Communication requirements include estimated bandwidth for the channels connecting the components. Additionally, information about the canonical ordering and dependencies between tasks that interact within the application will be documented here.

In addition to this basic functionality, we envisage three kinds of extended functionality: aggregation, dependency resolution, and storage, to which we turn next.

Free standing components and multi-component systems can themselves be combined into applications within an overarching formalism, such as may be the case where media conversion and encoding conversion are both required prior to other components becoming useful in the application context.

The broker must evaluate applications submitted to it for unresolved dependencies or incompatibilities. For example, the broker should be able to determine that the output of component A will be incompatible with input formats acceptable to component B, and intervene with a proposed solution (e.g. encoding conversion). This requires a formal dependency resolution mechanism.

Furthermore, the application specifications and result instances will be stored for later re-use either as static data sets or as services (where aggregation of components results in a new service likely to be commonly requested). In order to meet this requirement, the application specification must include sufficient information so as to comply with the discovery mechanism, namely metadata based on that standardized by the Open Language Archives Community. Through the use of linguistic metadata, including language technology extensions, we can fully describe the functionality of a resource such as on specified using the grid application specification language. Automated discovery will occur through the use of a metadata harvester based on OLAC's Viser system (Simons, 2002).

## 7.2    Architecture

The architectural design allows for a number of immediately identifiable extensions and implications which we consider for future research.

The first is the potential for a range of software architectures in NLE to independently generate application definitions, and hence leverage high performance computing services. This increases the potential application of computational grids to a wide variety of research contexts within natural language engineering whilst allowing researchers to focus on their domain of interest.

The second is a portal implementation that allows the user to select the data sources and types of analysis which are of interest and perform these transparently over a grid using the application specification to formalize their selections, and the broker as an intermediary. A more user-centric possibility is that a portal could be implemented where users of a lower technical knowledge could design applications us-ing a point and click interface to combine data and components into applications, and have these executed on a grid infrastructure, with the results readily available.

The third is that grid resource broker architectures can be extended to support to data-centric processing which will significantly impact on the application of grid services to natural language engineering as well as other disciplines which are data-intensive. Another area of extension is in the modality of grid implementations themselves, where natural language engineering's common requirement for large data sets is juxtaposed with typical grid implementations which assume that data is inherently efficiently transported across the network. The development of a data-centric broker is a distinct advantage to our field in this light.

The fourth is the vision that a number of entities will offer grid enabled natural language engineering services as participants in a virtual organization. Ideally these entities would provide a diverse range of data sources and grid services which could be discovered remotely, and combined easily, as well as providing raw computational resources. In promoting collaborative natural language engineering research, another possibility is for a computation grid "virtual organisation" to be established between cooperating research centres, where computationally intensive applications can be distributed over a domain specific grid infrastructure, with a range of tools already available, and with direct access to major data sources.

In addition to these largely infrastructural directions, entirely new components and applications within this framework can be developed. As an example, we earlier identified a 'Comparative Evaluation' application which would use Grid services to allow the user to build a large number of language models based on controlled, parameterized options and compare these. The power and scalability of a grid infrastructure allows us to do this efficiently, combined with ease of access through the architecture we describe here makes this an attractive applied field with potential for linguistic discovery.

As mentioned earlier, current grid processing models are largely processor-centric, yet natural language engineering data sources lend themselves more favorably towards data-centric implementations. Our research in NLP has spawned investigation of these paradigmatic constraints, and results in the development of a new version of NIMROD-G which is oriented towards data intensive, rather than parameter intensive computation.

## 8  Conclusion

In conclusion we can consider features which make our grid-based architecture attractive. The resource description model allows us to be highly descriptive in reference to data and processes, in addition to being transparently retrofitted to the wide variety of existing data and software. From a functional perspective, our model supports the aggregation of disparate data sources and components into a single application definition. Geographical distribution, platform dependencies and licensing arrangements are all included in the application's single XML document. Additionally, where various components have conflicting requirements (e.g. for input and output), we can resolve these dependencies. Issues such as storage requirements can be encoded directly into the application specification, allowing the broker to mitigate the optimal runtime environment for the application execution from the plethora of grid resources available. Extensions to the model allow for the intermediate results of various applications to be stored for later re-use. Our vision is that where existing partial analysis already exist, we can discover and re-use these rather than executing computationally in a repetitive fashion.

Computational grids enable the sharing and aggregation of geographically distributed resources for solving large-scale, resource and data intensive problems such as those found in natural language engineering. In this paper we have described a model for a software architecture which allows the ease of integration of such services with traditional natural language engineering development environments. The grid-enabled component-based architecture described in this paper is novel for a number of reasons.

Firstly, although the component-based model is not new, our motivation for choosing this model (namely, ease of interface with grid services) brings a new perspective to the overall design at both component and application levels. Not only does the model scale well, allowing larger, more complex and computationally intensive processes to be divided for more efficient processing, but it also allows for efficiency of execution at a single processor level. In turn, this may allow efficiencies to be gained where previously computational power, data storage or bandwidth imposed constraints.

Secondly, the use of a formal application specification is proposed, which allows our component-based model to be utilized directly as the input for a grid service broker, whilst allowing flexibility in the selection of an underlying framework. Additionally, applications can now be evaluated for optimality and interdependency prior to execution, thus allowing processing resources to be utilized more efficiently. A broker can observe that certain aggregated services are used repetitively (through detection of common threads in multiple application descriptions), and initiate the creation of discoverable services based on these selections. Previously collated applications can be re-used as complex components in later, larger applications, for example through the use of cached result sets for intermediate computation.

Thirdly we identify the need for existing grid service broker architectures to allow for data-centric processing, as opposed to currently implemented processor-centric approaches. The adoption of a broker based architecture and the open, accessible nature of the broker interface will allow a large class of natural language engineering applications to utilize grid services with corresponding gains in computational efficiency. Exploratory implementation in this area indicates that there is significant research required to adapt existing Grid technologies to facilitate data-intensive processing.

Finally the optional availability of previously compiled grid applications and output results (both result data sets and resultant services) are all made available as discoverable resources for future application design.

Our model enables researchers to leverage the power of the computational grid without acquiring an in-depth knowledge of the infrastructure at both design and implementation levels. It is our hope that a grid-based architecture for high-performance natural language processing will open the way for innovative research at both the software architecture and language engineering levels.

## Credits

The research described in this paper has been partially funded by the National Science Foundation under Grant Nos. 9910603 (International Standards in Language Engineering), 9978056, 9980009 (Talkbank) and by the Victorian Partnership in Advanced Computing Expertise Grant EPPNME092.2003.